\documentclass[12pt,twoside,a4paper]{article}
\usepackage{graphicx}

\def\lsim{\:\raisebox{-0.5ex}{$\stackrel{\textstyle<}{\sim}$}\:}

\begin{document}
\thispagestyle{empty} 

\title{
\vskip-3cm
{\baselineskip14pt
  \centerline{\normalsize DESY 12--013 \hfill ISSN 0418--9833}
  \centerline{\normalsize MZ-TH/12--07 \hfill} 
  \centerline{\normalsize LPSC 12019 \hfill} 
}
\vskip1.5cm
{\bf Inclusive Charmed-Meson Production}
\\[1.3ex]
{\bf at the CERN LHC}
\author{
  B.~A.~Kniehl$^1$, 
  G.~Kramer$^1$, 
  I.~Schienbein$^2$ 
  and H.~Spiesberger$^3$
  \vspace{2mm} \\
  \normalsize{
    $^1$ II. Institut f\"ur Theoretische
    Physik, Universit\"at Hamburg,
  }\\ 
  \normalsize{
    Luruper Chaussee 149, D-22761 Hamburg, Germany
  } \vspace{2mm}\\
  \normalsize{
    $^2$ Laboratoire de Physique Subatomique et de Cosmologie,
  } \\ 
  \normalsize{
    Universit\'e Joseph Fourier Grenoble 1,
  }\\
  \normalsize{
    CNRS/IN2P3, Institut National Polytechnique de Grenoble,
  }\\
  \normalsize{
    53 avenue des Martyrs, F-38026 Grenoble, France
  } \vspace{2mm}\\
  \normalsize{
    $^3$ Institut f\"ur Physik,
    Johannes-Gutenberg-Universit\"at,
  }\\ 
  \normalsize{
    Staudinger Weg 7, D-55099 Mainz, Germany
  } \vspace{2mm} \\
}
\normalsize{\date{}}
}

\maketitle
\begin{abstract}
\medskip
\noindent
We present predictions for the inclusive production of 
charmed hadrons at the CERN LHC in the general-mass 
variable-flavor-number scheme at next-to-leading order. 
Detailed numerical results are compared to data where 
available, or presented in a way to ease future 
comparisons with experimental results. We also point 
out that measurements at large rapidity have the 
potential to pin down models of intrinsic charm.
\end{abstract}


\clearpage

\section{Introduction}

Past measurements of charmed-hadron, $X_c$, production in 
high-energy scattering experiments \cite{Acosta:2003ax} have 
shown that inclusive heavy-quark production provides us with 
an interesting testing ground for the dynamics of the strong 
interaction. With the advent of high-statistics data from the 
LHC, we can expect that upcoming new results will allow us to 
test the predictions of QCD with much better precision. The 
LHC experiments will also extend into the region of higher 
transverse momenta, where theory is expected to provide more 
reliable predictions. This is not only important for testing 
QCD itself; but a good understanding of heavy-quark production 
is also vital for searches of new-physics phenomena.

Recent analyses of $D$-meson production at the LHC have been 
published by the ALICE \cite{alice:2011ka} and ATLAS 
collaborations \cite{atlas-note-2011}. Similar data that are 
being analyzed by the LHCb collaboration \cite{lhcb-confnote-2011} 
are expected to appear soon. The aim of the present article is 
to present predictions for the inclusive production of charmed 
hadrons at the LHC within the general-mass variable-flavor-number 
scheme (GM-VFNS). We extend results which have already been 
used by the experimental collaborations for comparisons 
with their data, and we discuss these results under a common 
perspective. Where possible we, therefore, use information 
from the experiments about the accessible phase space regions. 
In a recent paper \cite{Kniehl:2011bk}, we have considered 
the inclusive production of $B$ mesons, for which experimental 
results from the CMS collaboration are already available 
\cite{Khachatryan:2011mk,Chatrchyan:2011pw,Chatrchyan:2011vh}.

Theoretical predictions for the production of heavy-flavored 
mesons at high transverse momentum ($p_T$) are technically 
difficult to obtain due to the presence of two different 
scales, $p_T$ and the heavy-quark mass $m$. On the one hand, 
the heavy-quark mass can be considered as the large scale, 
since $m > \Lambda_{\rm QCD}$, making perturbative QCD 
applicable. When $m$ is the only large scale, as for example 
in the calculation of the total cross section or in 
predictions for $p_T$ distributions at small values of $p_T$, 
i.e.\ if $p_T$ is of the same order of magnitude as $m$, 
predictions from a fixed-order calculation are reliable. 
On the other hand, if the transverse momentum of the 
produced heavy quark is large compared with the heavy-quark 
mass, $p_T \gg m$, then large logarithms $\ln(p_T^2/m^2)$ 
have to be resummed.

In the first case, for $p_T \lsim m$, the traditional 
approach is called fixed-flavor-number scheme (FFNS) 
\cite{ffns-theory}. It is based on the assumption that 
the gluon and the light partons ($u, d, s$) are the only 
active partons. The heavy quark appears only in the final 
state, produced in the hard-scattering process of light 
partons. The heavy-quark mass is explicitly taken into 
account together with the transverse momentum  of the 
produced heavy meson assuming that $m$ and $p_T$ are of 
the same order. Predictions of the FFNS are reliable close 
to the threshold of heavy-quark production.

In the second case, for $p_T \gg m$, the large logarithmic 
terms have to be resummed. The well-known factorization 
theorem provides the foundation of this resummation by 
incorporating the large logarithms into parton distribution 
and fragmentation functions and imposing the DGLAP evolution 
equations. This approach requires that the heavy quark is treated 
as a parton, and as a consequence, one has to take into account 
additional processes where heavy quarks occur as incoming partons 
or by fragmentation from light partons. The heavy quark is 
treated as any other massless parton. If $m$ is neglected in 
the calculation of the hard-scattering matrix element, this 
approach is called zero-mass variable-flavor-number scheme 
(ZM-VFNS) \cite{zmvfns-theory}. The predictions in the ZM-VFNS 
are expected to be reliable only in the region of very large 
transverse momenta, since terms of the order of $m^2/p_T^2$ 
present in the hard-scattering cross sections are neglected.

In fact, it is not necessary to neglect $m$ altogether in a 
variable-flavor-number scheme. Instead, it is possible to 
absorb the large logarithms $\ln(p_T/m)$ into parton distribution 
and fragmentation functions, where they are resummed by imposing 
DGLAP evolution, as in the ZM-VFNS, while at the same time, 
$m$-dependent terms as obtained in the FFNS are retained in 
the hard-scattering cross sections. This so-called general-mass 
variable-flavor-number scheme (GM-VFNS) thus combines the virtues 
of the FFNS and the ZM-VFNS. The required subtraction of 
logarithmic terms, which are related to initial- and final-state 
singularities, can be defined using the usual $\overline{\rm MS}$ 
prescription, which guarantees the universality of parton 
distribution and fragmentation functions and allows for a 
meaningful comparison of data from different measurements.
This approach has been applied in the calculation of charmed-meson 
production in $p\bar{p}$ collisions in Ref.\ \cite{KKSS} 
and compared to data from the CDF collaboration 
\cite{Acosta:2003ax} at the Fermilab Tevatron. 

\section{Setup and Input}

The theoretical background and explicit analytic results of the 
GM-VFNS approach have been discussed in detail previously in 
Ref.\ \cite{KKSS} and the references cited therein. Here we only 
describe our choice of input for the present numerical analysis.

Throughout we use as parton distribution functions (PDFs) the 
set CTEQ6.6 \cite{CTEQ6.6} as implemented in LHAPDF 
\cite{LHAPDF}, except where we discuss uncertainties related 
to the choice of the PDFs. The fragmentation functions (FFs) 
determined in Ref.~\cite{Kneesch:2007ey} are used wherever 
possible, i.e.\ for the production of $D^0$, $D^{\pm}$, and 
$D^{\ast\pm}$ mesons. They are based on fits to the presently 
most precise data on charmed-meson production from the CLEO 
collaboration \cite{Artuso:2004pj} at LEPP CESR and from the 
Belle collaboration \cite{Seuster:2005tr} at the KEK collider 
for $B$ physics (KEKB). These FFs always refer to the average 
over charge-conjugated states, and our results below have to be 
understood as averaged cross sections $(\sigma(D) + 
\sigma(\overline{D}))/2$. For $D_s$ and $\Lambda_c$ production, 
we have to resort to the earlier determination of FFs described in 
Ref.~\cite{Kniehl:2006mw}. These FFs were determined by fitting 
the fractional-energy spectra of the $X_c$ hadrons measured by 
the OPAL collaboration \cite{Alexander:1996wy,Ackerstaff:1997ki} 
in $e^+e^-$ annihilation on the $Z$-boson resonance at the CERN 
LEP1 collider. These data have rather large experimental errors 
and the disadvantage of being at the rather large scale of 
the $Z$-boson mass, far away from the typical scales of $X_c$ 
production presently observed at the LHC.

The subtractions related to renormalization as well as to the 
factorization of initial- and final-state singularities requires 
the introduction of scale parameters $\mu_R$, $\mu_I$, and 
$\mu_F$. We fix these scales by the transverse mass of the 
produced charm quark, $m_T = \sqrt{p_T^2 + m^2}$. In order 
to exploit the freedom in the choice of scales, we introduce 
scale parameters $\xi_i$ ($i = R$, $I$, $F$) by setting $\mu_i 
= \xi_i m_T$. Error bands describing the theoretical 
uncertainties can then be determined by varying the values of 
$\xi_i$ independently by factors of two up and 
down while keeping any ratio of the $\xi_i$ smaller than or 
equal to two. 

We emphasize that the uncertainties related to scale 
variations are by far the dominating source of theoretical 
uncertainties. For PDF-related uncertainties, we will show 
explicit results below. For the influence of variations of 
the value of the charm-quark mass we refer to our previous 
work \cite{KKSS}. There we have shown that those terms that 
carry a dependence on $m$, coming from the hard-scattering 
matrix element, are small. Therefore, we always keep $m=1.5$ 
GeV in the present work.

\section{Numerical Results}

\begin{figure}[b!]
\begin{center}
\includegraphics[scale=0.9]{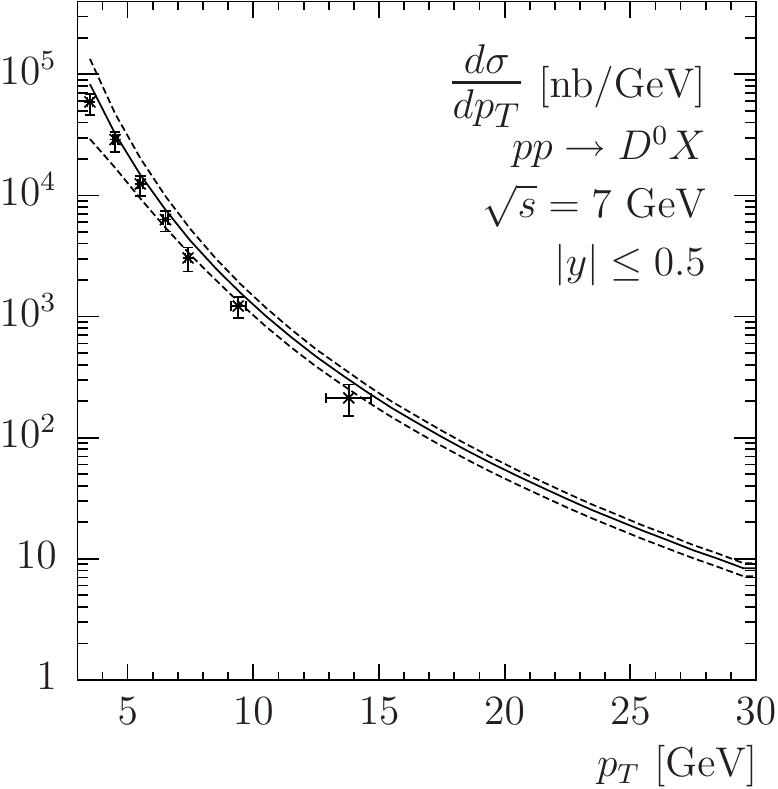}
~~~~~
\includegraphics[scale=0.9]{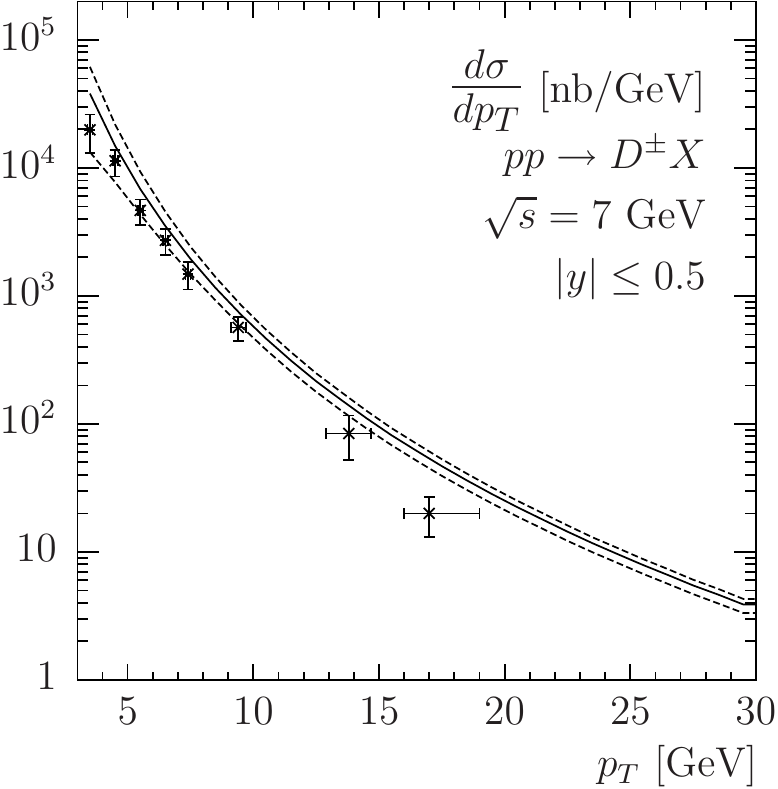}
\\
\includegraphics[scale=0.9]{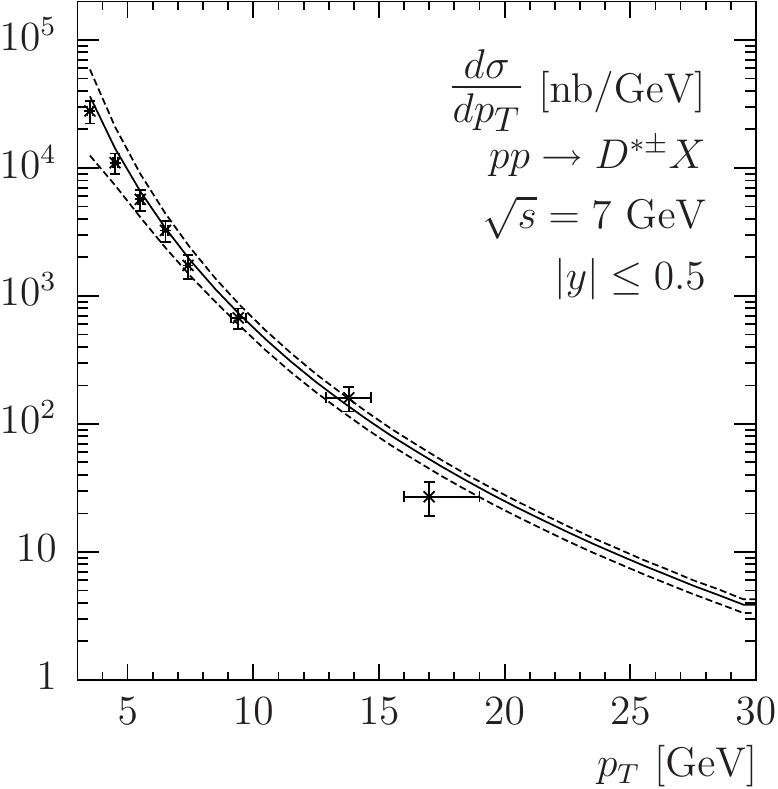}
\end{center}
\caption{\label{fig:1}
Differential cross section $\mathrm{d}\sigma/\mathrm{d}p_T$ 
as a function of $p_T$ for $p+p \to D+X$ with (a) $D=D^0$, 
(b) $D=D^{\pm}$, and (c) $D=D^{\ast\pm}$ integrated over 
rapidity in the range $-0.5 \leq y \leq 0.5$ for $\sqrt{s}=7$~TeV 
at NLO in the GM-VFNS using the FFs of Ref.~\cite{Kneesch:2007ey} 
and the CTEQ6.6 PDFs. The full lines are obtained for the 
default choice of scale parameters $\xi_R = \xi_I = \xi_{F} 
= 1$, and the error band (dashed lines) are from independent 
variations of $\xi_R$, $\xi_I$, and $\xi_{F}$ as described in 
the text. The data points are taken from Ref.\ \cite{alice:2011ka}. 
}
\end{figure}

\begin{figure}[b!]
\begin{center}
\includegraphics[scale=0.9]{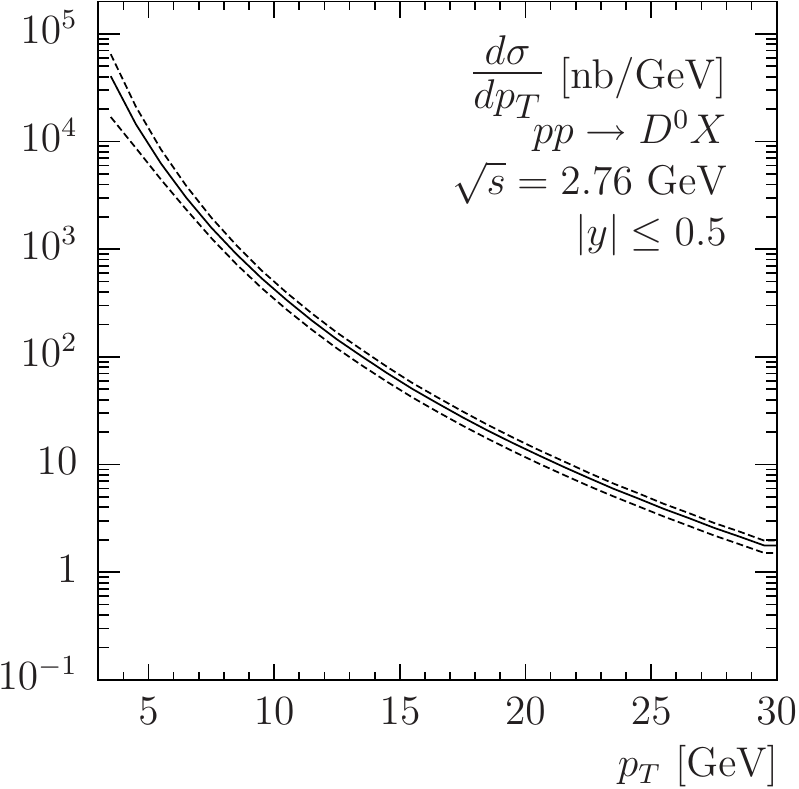}
~~~~~
\includegraphics[scale=0.9]{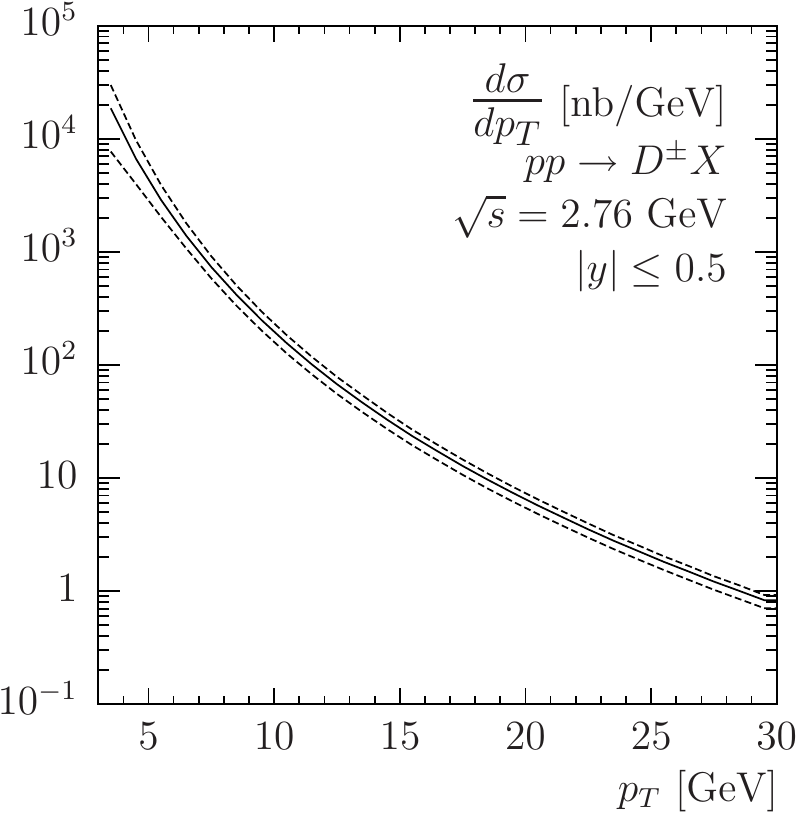}
\\
\includegraphics[scale=0.9]{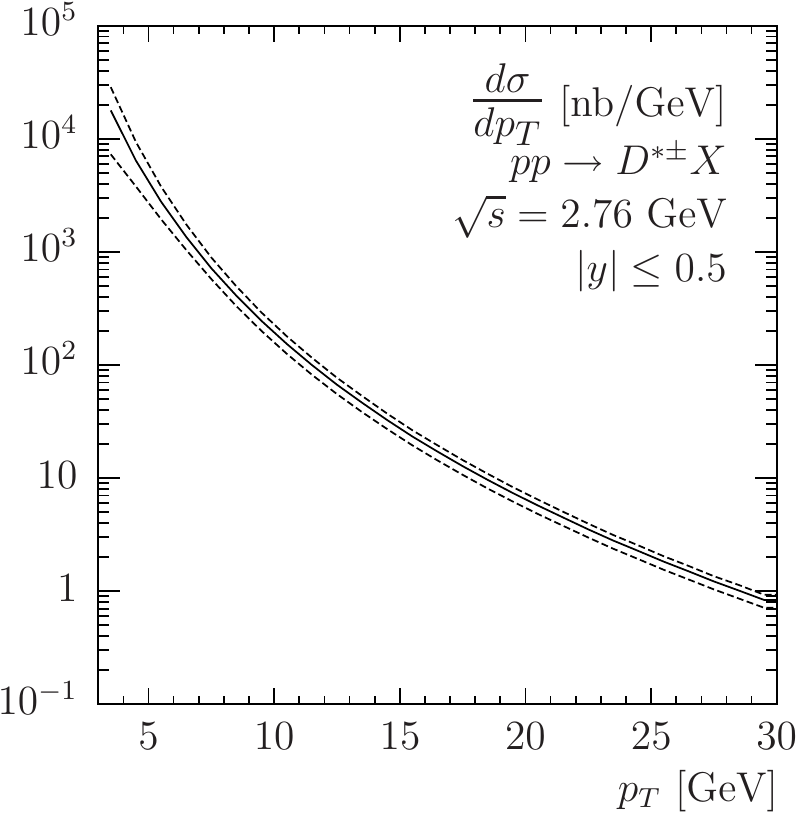}
\end{center}
\caption{\label{fig:2}
Same as Fig.\ \ref{fig:1} but for $\sqrt{s}=2.76$~TeV. 
} 
\end{figure}

The ALICE collaboration has published results for the 
differential cross section $d\sigma / dp_T$ at $\sqrt{s} 
= 7$ TeV in the central rapidity range $|y| \leq 0.5$. 
In Fig.\ \ref{fig:1}, we show the predictions of the GM-VFNS 
for $p_T \geq 3$ GeV, where we identify $y$ with the 
pseudorapidity of the produced heavy meson. We include a 
comparison with the data points of ALICE taken from Ref.\ 
\cite{alice:2011ka} for the production of $(D^0 + 
\overline{D}^0)/2$, $(D^+ + D^-)/2$, and $(D^{\ast +} + 
D^{\ast -})/2$. An error band due to scale variations as 
described in the preceding section is also shown. The 
agreement between theory and experiment is good except at 
the largest values of $p_T$, where the data lie somewhat 
below the error band. Data for the smaller center-of-mass 
energy $\sqrt{s} = 2.76$ TeV, also for $|y| \leq 0.5$, are 
currently under investigation. We, therefore, present results 
for this forthcoming analysis in Fig.\ \ref{fig:2} using the 
same conventions as in Fig.\ \ref{fig:1}.

\begin{figure}[b!]
\begin{center}
\begin{tabular}{cc}
\parbox{0.5\textwidth}{
\includegraphics[scale=0.9]{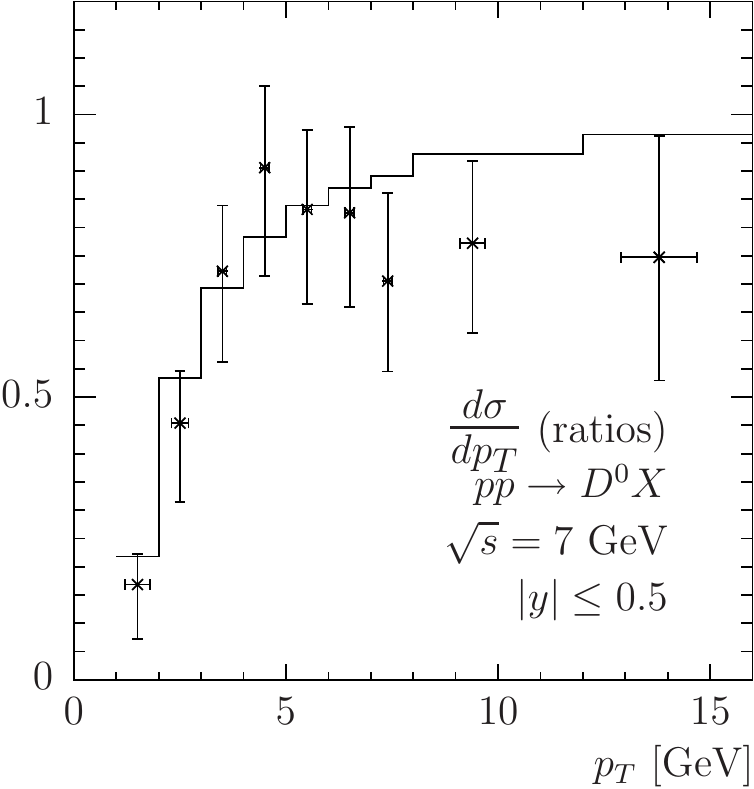}
}
&
\parbox{0.5\textwidth}{
\includegraphics[scale=0.9]{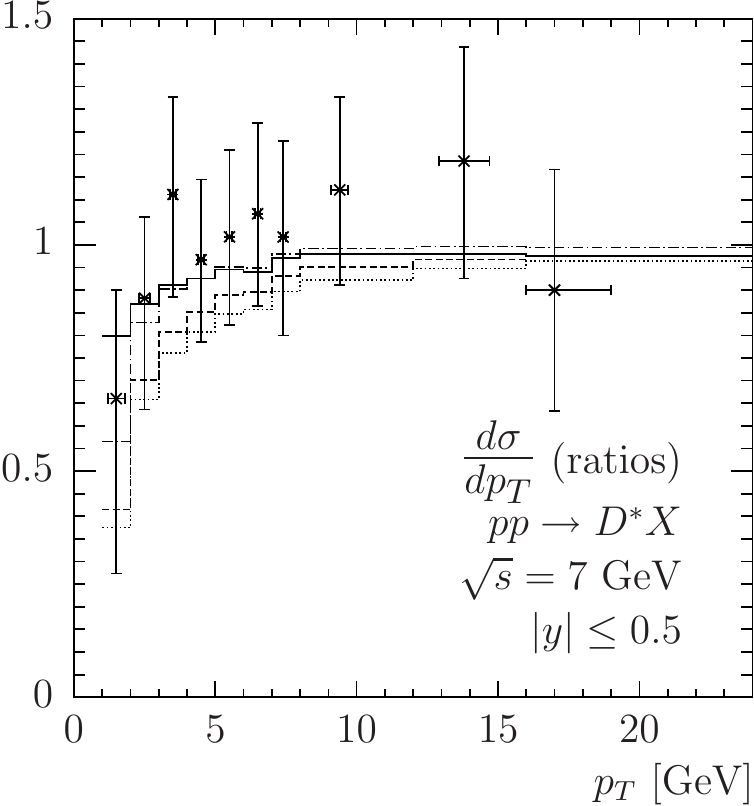}
}
\\
(a) & (b) 
\end{tabular}
\end{center}
\caption{\label{fig:3}
(a) Ratios of $d\sigma/dp_T$ for $D^0$-meson production at 
ALICE at $\sqrt{s}=7$~TeV. All cross sections are normalized 
to the GM-VFNS prediction with $\xi_i=1$. Data points, also 
normalized to the GM-VFNS calculation, are taken from Ref.\ 
\cite{alice:2011ka}. The histogram is obtained using 
$\xi_I = \xi_F = 0.7$ and $\xi_R = 1$.
(b) Ratios of $d\sigma/dp_T$ for $D^{\ast}$-meson production at 
ALICE at $\sqrt{s}=7$~TeV for different PDFs. All cross sections 
are normalized to the GM-VFNS prediction with $\xi_I = \xi_F = 
0.7$. Data points are taken from Ref.\ \cite{alice:2011ka} and 
normalized to the GM-VFNS calculation with $\xi_I = \xi_F = 0.7$. 
Full histogram: CT10 \cite{Lai:2010vv}, 
dashed: MSTW08-NLO \cite{Martin:2009iq}, 
dotted: NNPDF 2.1 \cite{Ball:2011mu},
dash-dotted: HERAPDF 1.5 (NLO) \cite{herapdf15}. 
} 
\end{figure}

At low transverse momentum, $p_T \lsim 3$ GeV, scale uncertainties 
are large and amount to about $\pm 65\,\%$; they decrease to 
values below $+30\, \%$ $(-15\,\%)$ at $p_T \simeq 30$ GeV. In 
Fig.\ \ref{fig:3}a, we show the cross section ratios for $d\sigma 
/ dp_T$ normalized to the GM-VFNS default prediction with 
$\xi_i = 1$. Also the data points of the ALICE experiment are 
normalized to the GM-VFNS result with $\xi_i = 1$. At 
$p_T \lsim 3$, the  data are much smaller than the default 
prediction. However, it turns out that a very good description 
of the data also in the first three $p_T$ bins can be obtained 
by choosing $\xi_I = \xi_F = 0.7$ (see the full histogram). 
This choice of scales corresponds to setting 
$\mu_i = \sqrt{m_T^2/2}$. 

We repeat that the uncertainties of theoretical predictions 
due to scale variations are dominating. This can be seen 
by comparing the results shown in Fig.\ \ref{fig:3}a with 
those in Fig.\ \ref{fig:3}b, where we present ratios of 
$d\sigma / dp_T$ for different choices of the PDF parametrizations 
with $\xi_I = \xi_F = 0.7$. All results including the data 
are normalized to the default evaluation using the PDF set 
CTEQ6.6. We notice that all PDFs reproduce the data very well 
inside the experimental errors except in the first $p_T$ bin, 
where the default prediction, corresponding to unit ratio, 
lies outside the error. All other PDF choices, i.e.\ MSTW08-NLO 
\cite{Martin:2009iq}, NNPDF2.1 \cite{Ball:2011mu}, HERPDF1.5-NLO 
\cite{herapdf15}, and CT10 \cite{Lai:2010vv}, yield predictions 
inside the experimental errors. 

We should emphasize here that, with our choice of scale 
parameters, $\mu_I$ and $\mu_F$ can fall below $m$ at the 
lowest $p_T$ values. In our implementation, we freeze the 
scales at $\mu_I = \mu_F = m = 1.5$ GeV, so that the PDFs 
and FFs do not evaluate to zero. The observed strong 
suppression in the lowest $p_T$ bin in Fig.\ \ref{fig:3}b is 
then partly due to the fact that the value of $m$ used in the 
more recent PDF fits is $m = 1.4$ GeV, and not $m = 1.3$ GeV 
as in the case of CTEQ6.6. As a consequence, the charm PDF of 
set CTEQ6.6 is larger due to the longer evolution path as 
compared to MSTW08-NLO, NNPDF2.1, and HERAPDF1.5-NLO.

Actually, due to the different values of $m$ used in the PDF 
fits, there is some residual $m$ dependence of the predicted 
cross sections at low values of $p_T$. The value $m = 1.5$ GeV 
used in our calculation agrees with the one in the FF fits from 
Ref.\ \cite{Kneesch:2007ey}, but not with the one in the PDFs 
used here. For example, CTEQ6.6 and CT10 use $m = 1.3$ GeV while 
in the parametrizations MSTW08-NLO, NNPDF 2.1, and HERAPDF 1.5 
(NLO) $m = 1.4$ GeV is adopted. For consistency, one should use 
the same value of $m$ in all three components of the cross 
section formula, PDFs, FFs and hard-scattering matrix elements. 
This would, however, require separate fits of the FFs as 
functions of $m$, a task which is left for future studies. 
Similarly, it would not completely be consistent to use different 
values of $\alpha_s$ in the partonic cross sections and in the 
PDFs and FFs. For our default choice of PDFs, CTEQ6.6, as well 
as for CT10, our choice $\alpha_s(M_Z^2) = 0.118$ agrees with 
what was used in the PDF and FF fits. For MSTW08-NLO and NNPDF 
2.1, the value of $\alpha_s$ is only slightly larger than for 
CTEQ6.6, and for HERAPDF 1.5 (NLO) the value $\alpha_s(M_Z^2) 
= 0.1176$ was used, which is nearly the same as for CTEQ6.6. 
We are thus confident that our evaluations are consistent with 
respect to the choice of $\alpha_s$. 

\begin{figure}[h!]
\begin{center}
\includegraphics[scale=0.9]{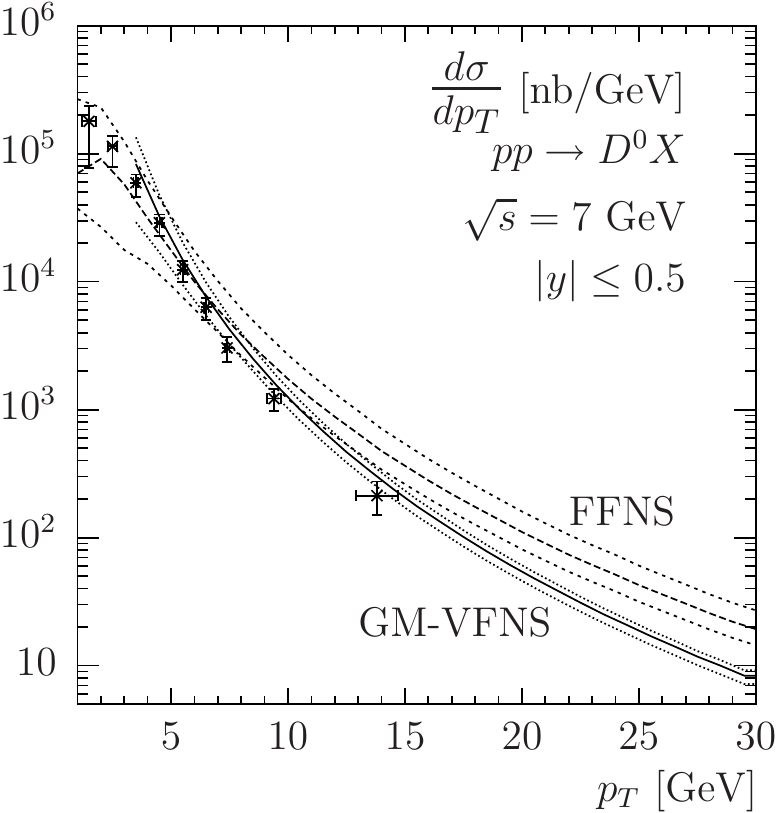}
\end{center}
\caption{\label{fig:4}
$p_T$ distribution $\mathrm{d}\sigma/\mathrm{d}p_T$ for $p+p 
\to D^0 + X$ integrated over rapidity in the range $|y| \leq 
0.5$ for $\sqrt{s} = 7.0$~TeV at NLO in the GM-VFNS (full line) 
as in Fig.\ \ref{fig:1}, compared with the results of the FFNS 
(dashed line). Dotted lines describe the corresponding error 
bands from scale variations as described in the text. The 
points with error bars are data from the ALICE collaboration 
\cite{alice:2011ka}.
} 
\end{figure}

Although the GM-VFNS calculation is able to describe the data 
at the lowest values of $p_T$ if the factorization scales are 
fixed by choosing $\xi_I = \xi_F = 0.7$, one cannot expect a 
fully satisfactory description for $p_T \rightarrow 0$. This is 
due to contributions with a charm quark in the initial state, 
where a massless approach is required. In the low-$p_T$ range, 
one expects the FFNS approach to be more reliable. A comparison 
of the predictions of the GM-VFNS and the FFNS for the case 
of $D^0$ production, using $\xi_i = 1$, is shown in Fig.\ 
\ref{fig:4}. The FFNS calculation is done without including 
a FF; the transition from the charm quark to the charmed meson 
is, however, taken into account by multiplying the parton-level 
results with the branching ratio $BR(c\rightarrow D^0) = 0.62$. 
One can observe that the FFNS describes the turn-over towards 
the production threshold at low values of $p_T$ correctly, but 
it fails to describe the data in the high-$p_T$ tail. This 
demonstrates the importance of resumming terms involving the 
large collinear logarithm, $\ln (p_T/m)$, which is done in the 
GM-VFNS, but not in the FFNS. The resummation is also responsible 
for the considerable reduction of the size of the error band from 
scale variations. Taking into account a scale-independent FF may 
improve the agreement between the FFNS and the data at high 
values of $p_T$, but the theory loses its predictability here, 
since the FF can not be defined in a universal way in the FFNS. 
From Fig.\ \ref{fig:4}, one can also read off that the curves 
from the GM-VFNS and the FFNS meet each other at about $p_T 
\simeq 5$ GeV. This would be the $p_T$-value where the 
theoretical prescription should switch from a fixed to a 
variable-flavor-number scheme. We do not follow this 
possibility here, but leave a study of the matching between 
the GM-VFNS and the FFNS to future work.

\clearpage


\begin{figure}[b!]
\begin{center}
\includegraphics[scale=0.8]{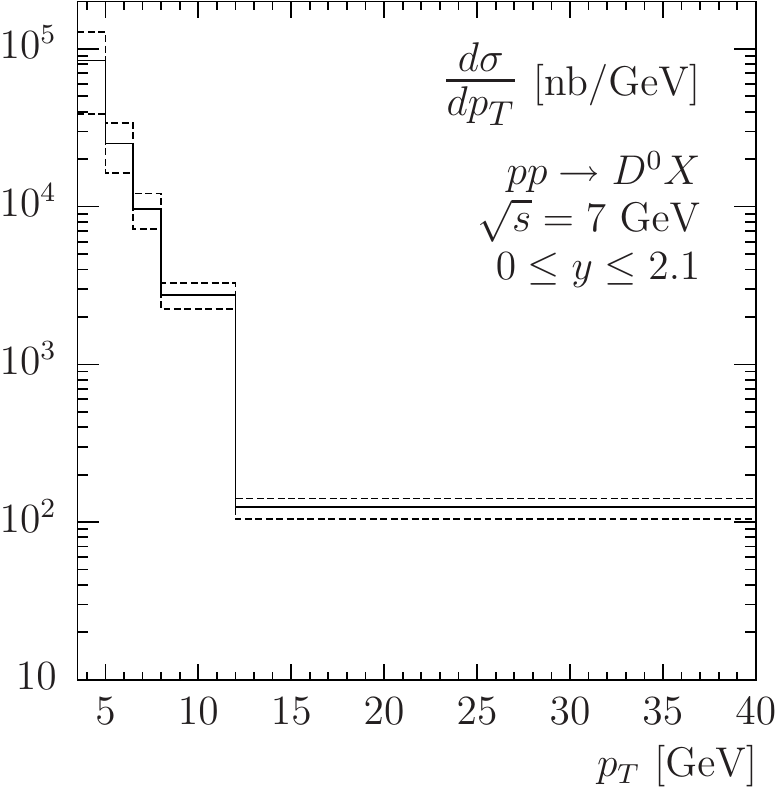}
~~~~~
\includegraphics[scale=0.8]{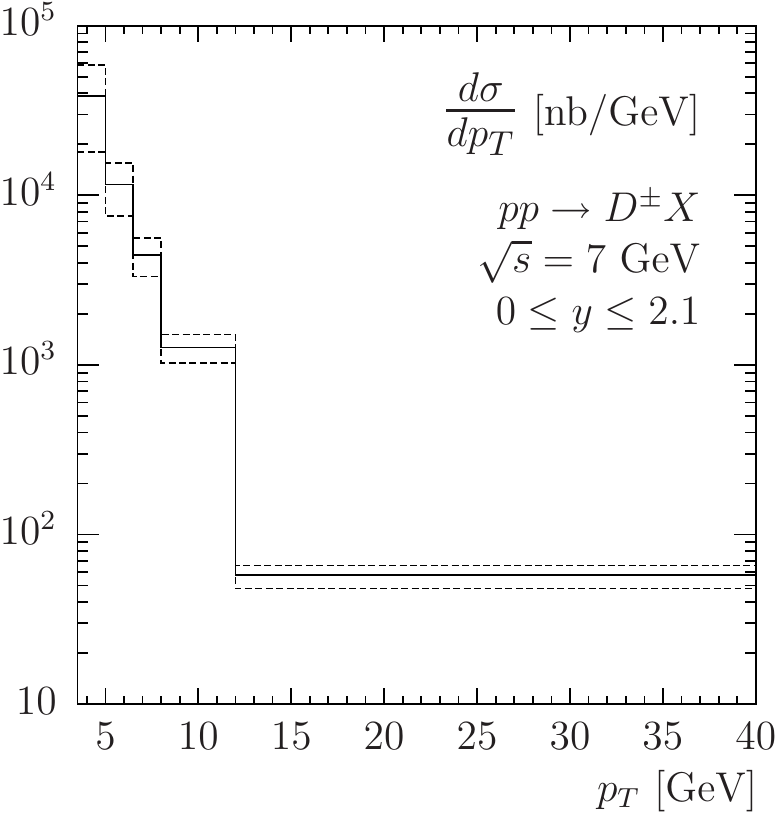}
\\
\includegraphics[scale=0.8]{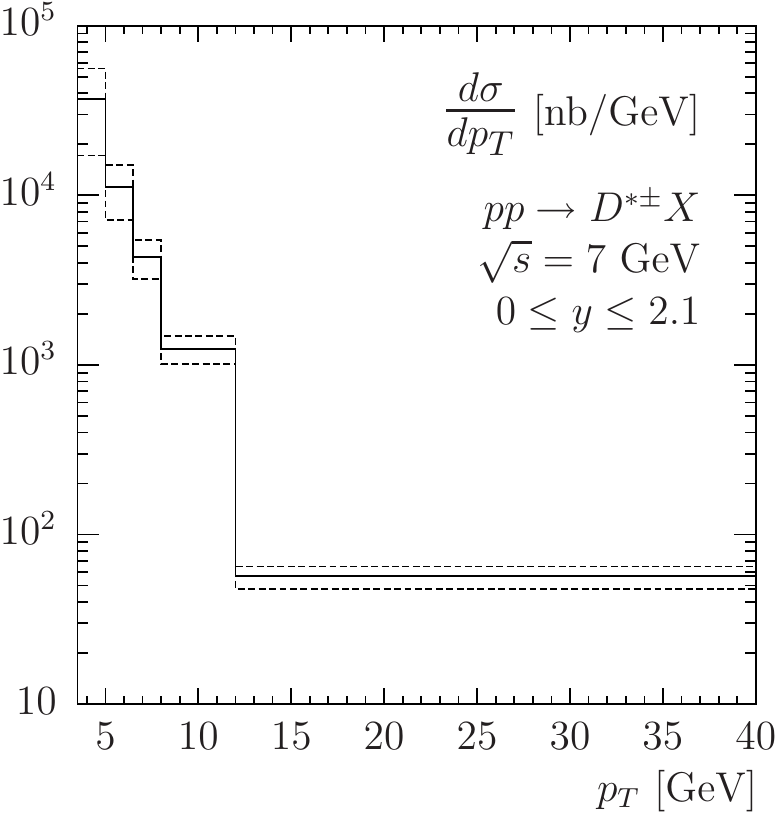}
\end{center}
\caption{\label{fig:5}
$p_T$ distributions $\mathrm{d}\sigma/\mathrm{d}p_T$ of $p+p 
\to D + X$ with (a) $D = D^0$, (b) $D = D^{\pm}$, and (c) $D = 
D^{\ast\pm}$ integrated over rapidity in the range $0 \leq y 
\leq 2.1$ for $\sqrt{s} = 7.0$~TeV at NLO in the GM-VFNS using 
the FFs of Ref.~\cite{Kneesch:2007ey} and the CTEQ6.6 PDFs. The 
full histograms are obtained for the default choice of scale 
parameters, $\xi_R = \xi_I = \xi_F = 1$, and the error bands 
(dashed histograms) are from independent variations of $\xi_R$, 
$\xi_I$, and $\xi_F$ as described in the text. 
} 
\end{figure}

\begin{figure}[b!]
\begin{center}
\includegraphics[scale=0.8]{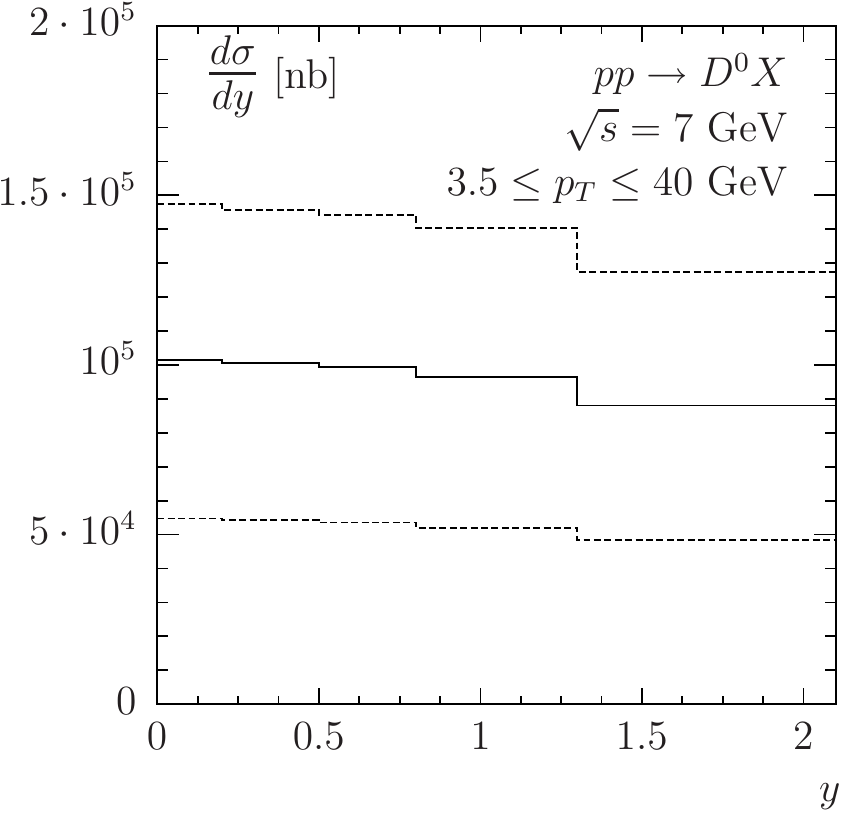}
~~~~~
\includegraphics[scale=0.8]{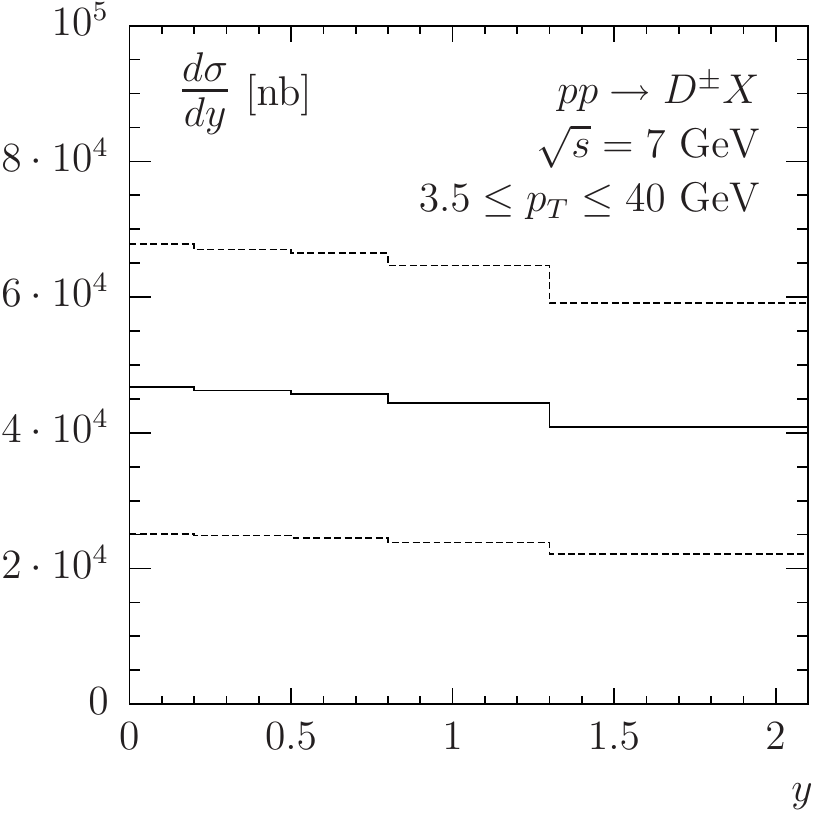}
\\
\includegraphics[scale=0.8]{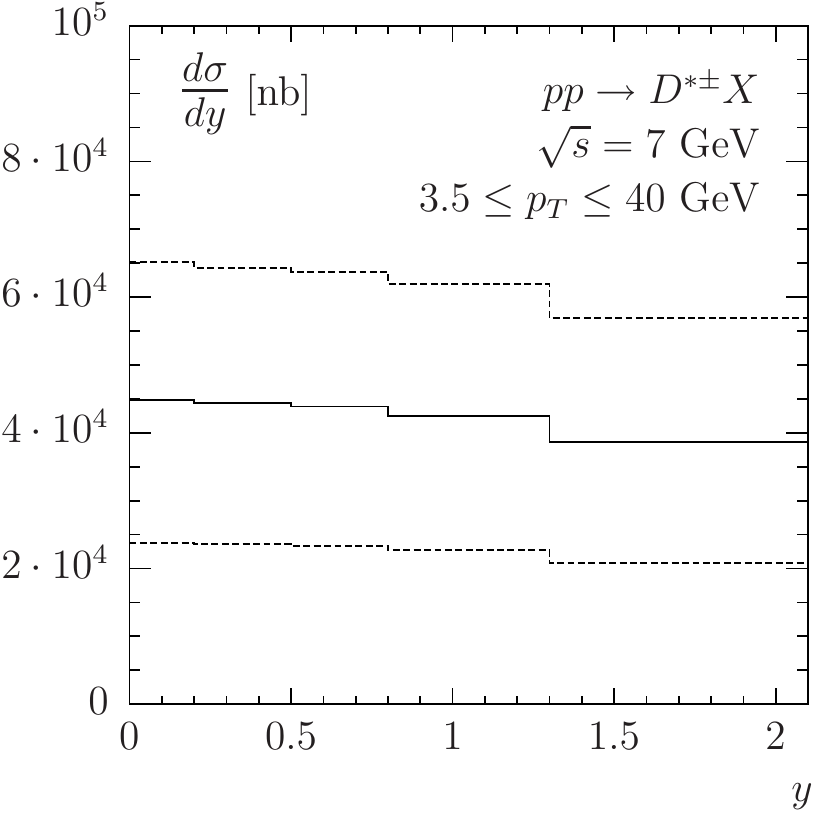}
\end{center}
\caption{\label{fig:6}
Rapidity distributions $\mathrm{d}\sigma/\mathrm{d}y$ of $p+p 
\to D + X$ with (a) $D = D^0$, (b) $D = D^{\pm}$, and (c) $D = 
D^{\ast\pm}$ integrated over transverse momentum in the range 
3.5 GeV $\leq p_T \leq$ 40 GeV for $\sqrt{s} = 7.0$~TeV at NLO 
in the GM-VFNS using the FFs of Ref.~\cite{Kneesch:2007ey}. The 
full histograms are obtained for the default choice of scale 
parameters, $\xi_R = \xi_I = \xi_F = 1$, and the error bands 
(dashed histograms) are from independent variations of $\xi_R$, 
$\xi_I$, and $\xi_F$ as described in the text. 
} 
\end{figure}

Now we present results for a comparison with experimental data 
from the ATLAS collaboration following the analysis described 
in Ref.\ \cite{atlas-note-2011}. The corresponding measurement 
covers a larger $p_T$ range, up to $p_T = 40$ GeV; also the $y$ 
range is wider and extends up to $y=2.1$. We choose the bins in 
$p_T$ and $y$ as in Ref.~\cite{atlas-note-2011}. Figures 
\ref{fig:5} and \ref{fig:6} show results for $d\sigma/dp_T$ and 
$d\sigma/dy$, respectively, for $D^0$, $D^{\pm}$, and 
$D^{\ast}$-meson production. Again, these results are for the 
average of the charge-conjugated states. We also remark that an 
experimentally relevant contribution coming from the 
fragmentation of $b$ quarks into $D$ mesons is not taken into 
account in our theoretical predictions. The behavior of these 
differential cross sections is similar to what we found above 
for the case of the ALICE data, i.e.\ we observe large scale 
variations at low values of $p_T$. The comparison with the ATLAS 
data as shown in Ref.\ \cite{atlas-note-2011} reveals that our 
calculations are in good agreement with the data within the 
observed large uncertainties. 

\clearpage


\begin{figure}[b!]
\begin{center}
\begin{tabular}{cc}
\parbox{0.5\textwidth}{
\includegraphics[scale=0.48]{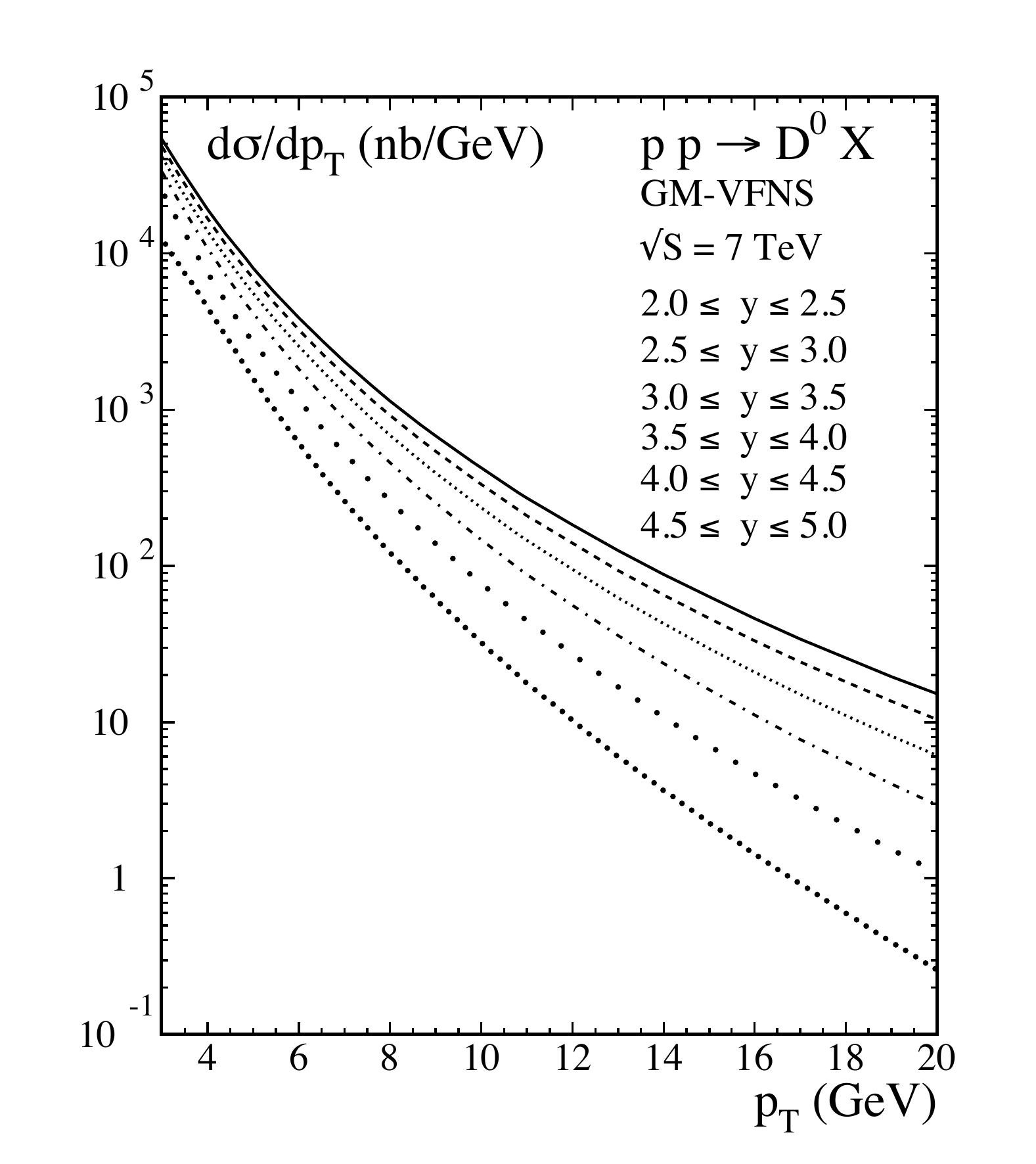}
}
&
\parbox{0.5\textwidth}{
\includegraphics[scale=0.48]{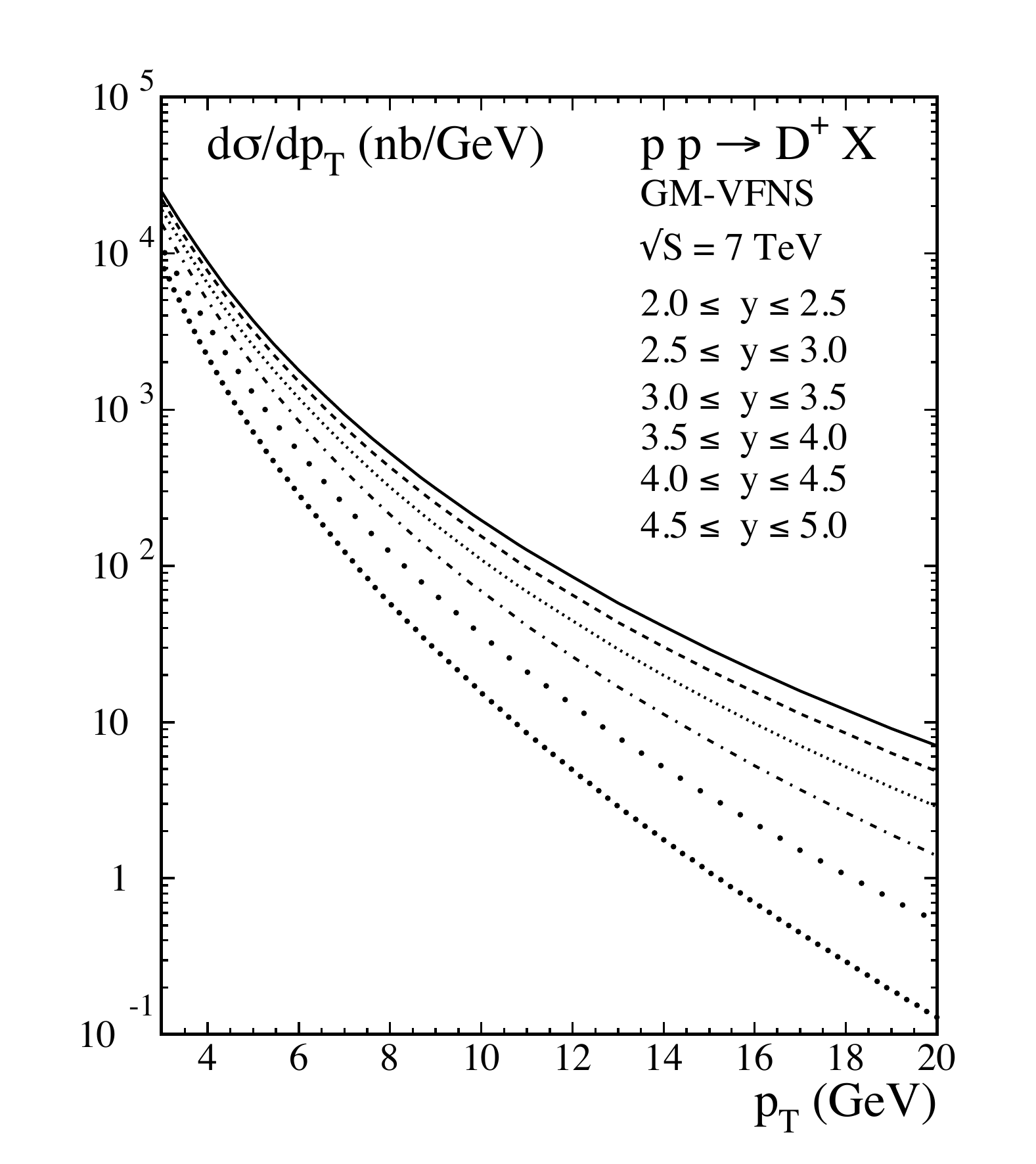}
}
\\[-1ex]
(a) & (b)
\\
\parbox{0.5\textwidth}{
\includegraphics[scale=0.48]{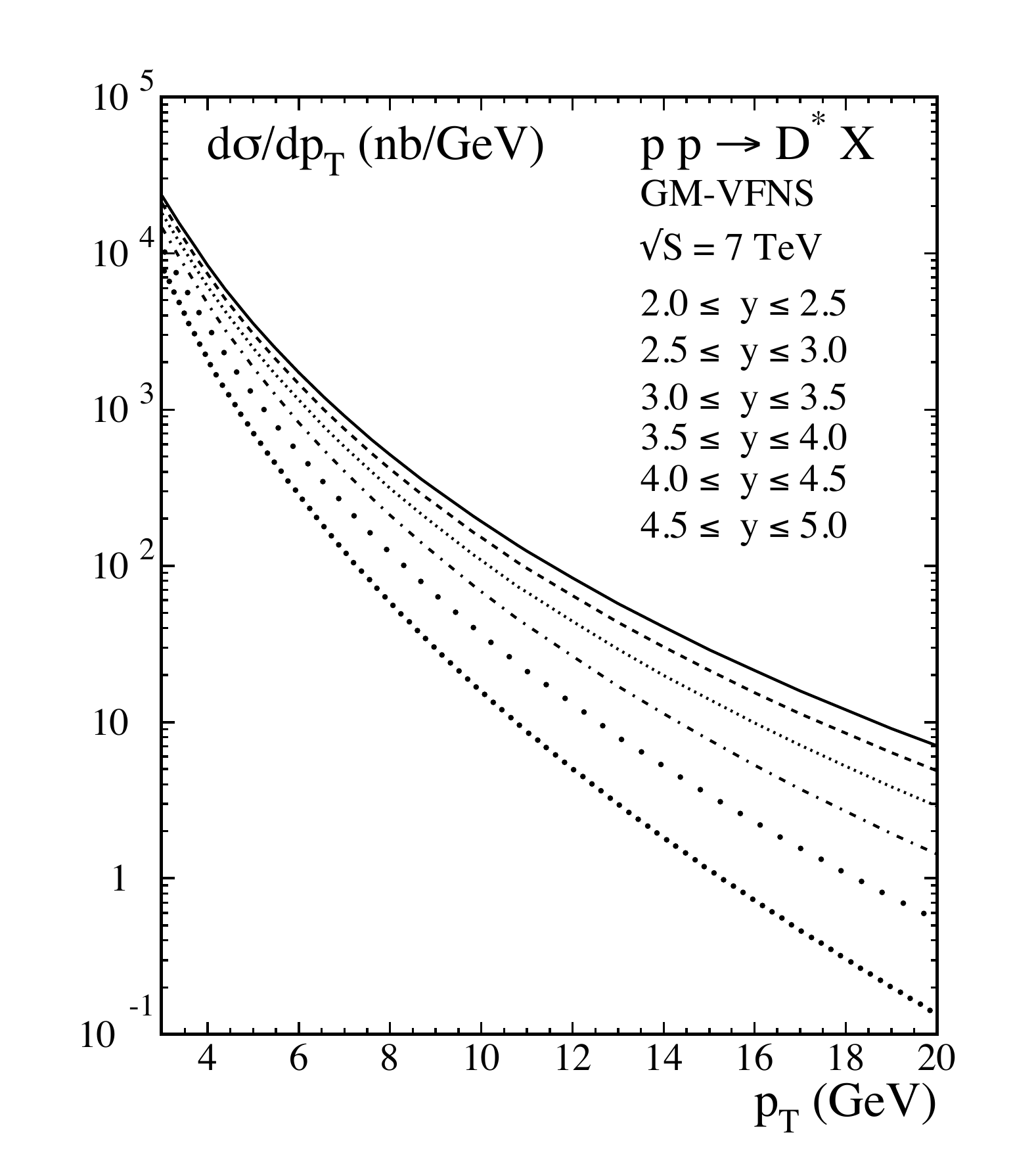}
}
&
\parbox{0.5\textwidth}{
\includegraphics[scale=0.48]{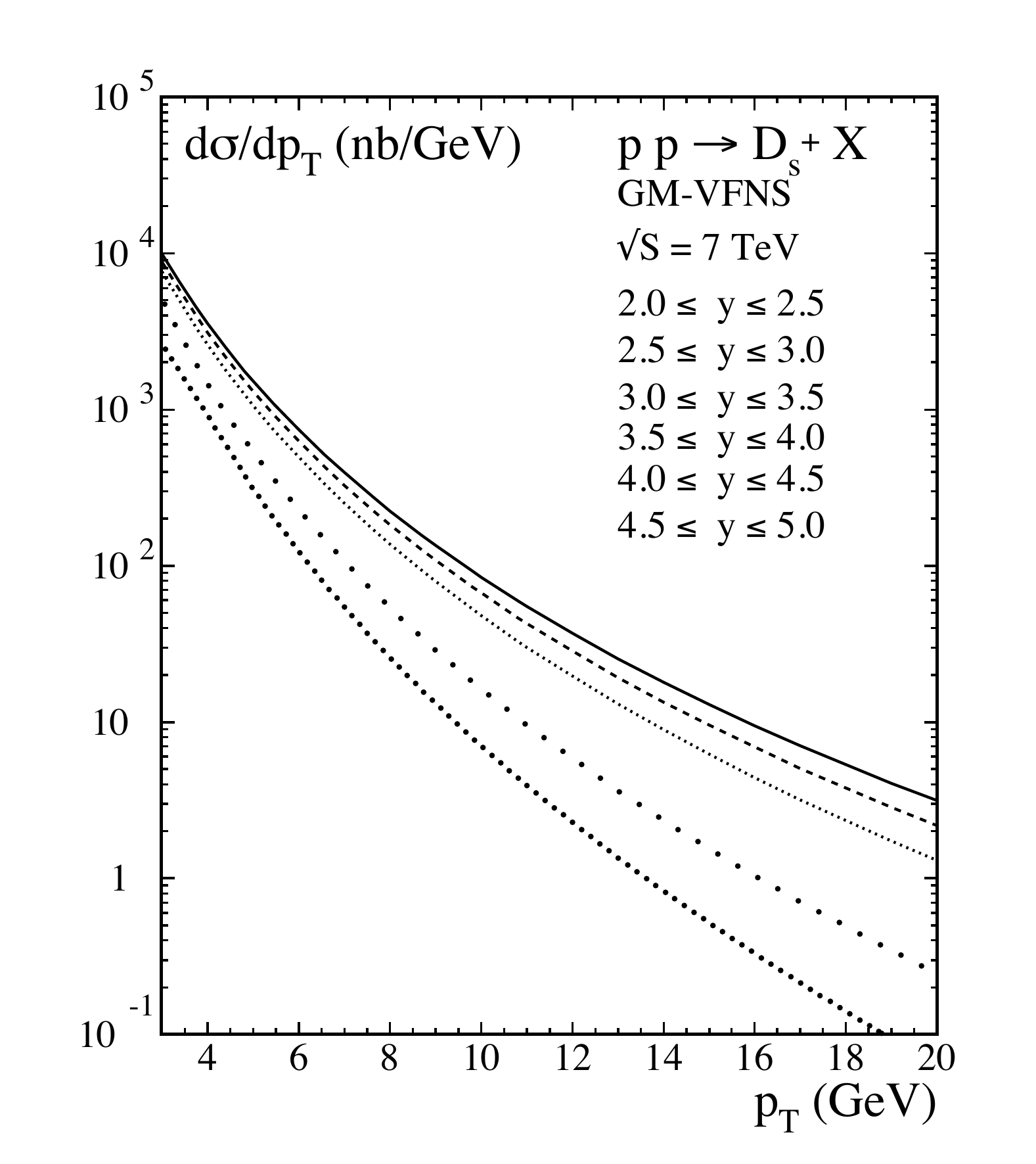}
}
\\[-1ex]
(c) & (d) 
\end{tabular}
\end{center}
\caption{\label{fig:7}
$p_T$ distributions $\mathrm{d}\sigma/\mathrm{d}p_T$ of $p+p \to 
D+X$ with (a) $D=D^0$, (b) $D=D^{\pm}$, (c) $D=D^{\ast\pm}$, and 
(d) $D=D_s^{\pm}$ for $\sqrt{s} = 7$~TeV at NLO in the GM-VFNS 
using the FFs of Ref.~\cite{Kneesch:2007ey} for $D^0$, $D^{\pm}$, 
and $D^{\ast\pm}$ and the FFs of Ref.~\cite{Kniehl:2006mw} for 
$D_s^{\pm}$. The various lines represent the default predictions 
for $\xi_R = \xi_I = \xi_F=1$, integrated over the rapidity 
regions indicated in the figures (larger rapidities correspond 
to smaller cross sections).
} 
\end{figure}

\begin{figure}[b!]
\begin{center}
\includegraphics[scale=0.5]{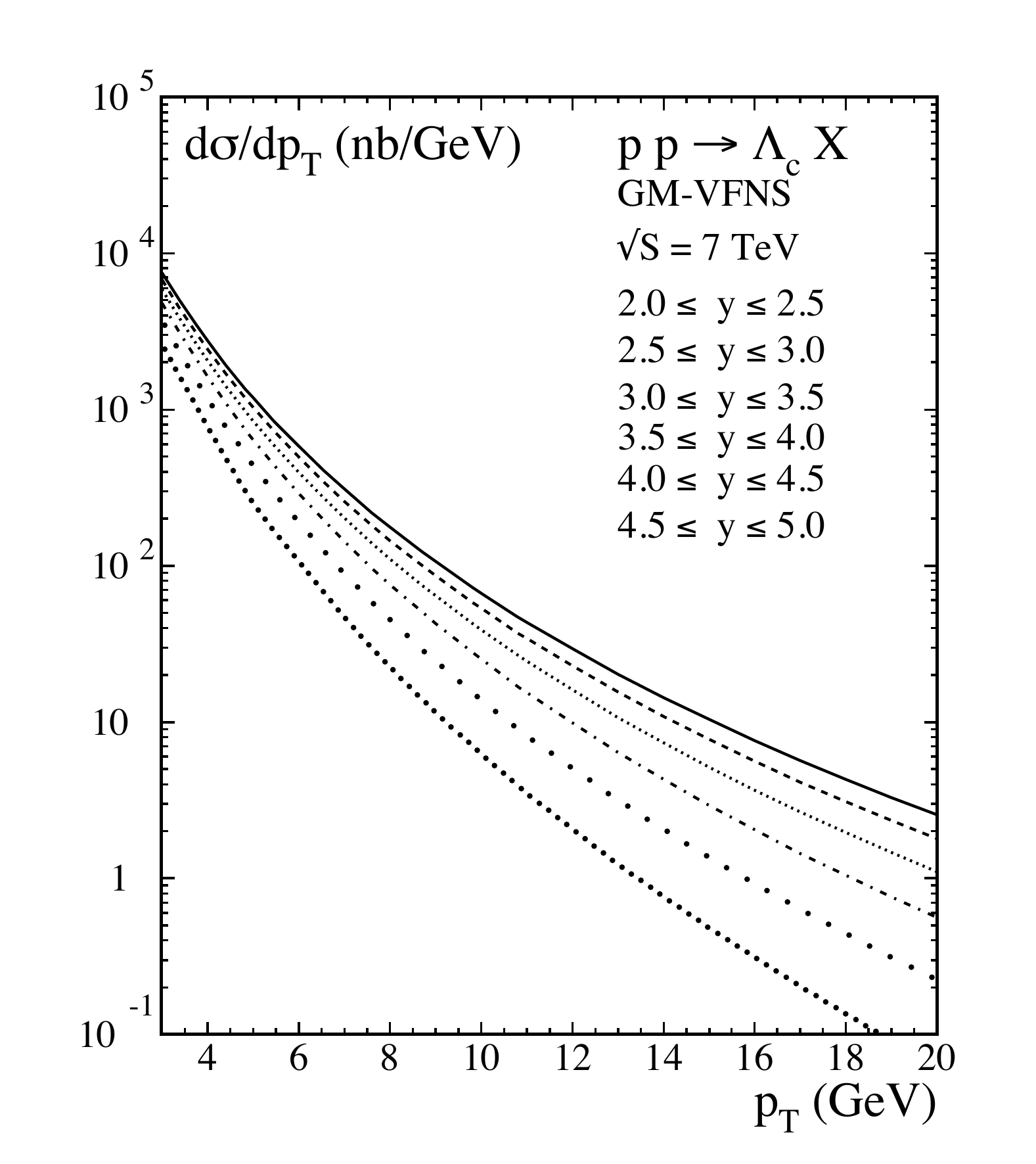}
\end{center}
\caption{\label{fig:8}
$p_T$ distributions $\mathrm{d}\sigma/\mathrm{d}p_T$ of $p+p \to 
\Lambda_c + X$ for $\sqrt{s} = 7$~TeV at NLO in the GM-VFNS 
using the FFs of Ref.~\cite{Kniehl:2006mw}. The various lines 
represent the default predictions for $\xi_R = \xi_I = \xi_F = 
1$, integrated over the rapidity regions indicated in the figure 
(larger rapidities correspond to smaller cross sections).
} 
\end{figure}

\begin{figure}[b!]
\begin{center}
\includegraphics[scale=0.72]{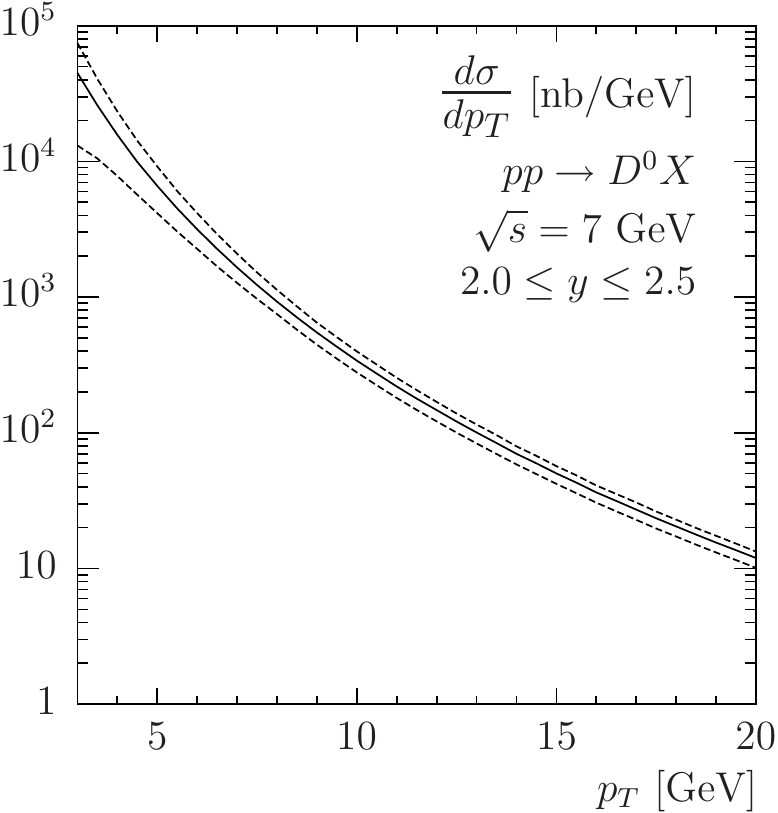}
~~~~~
\includegraphics[scale=0.72]{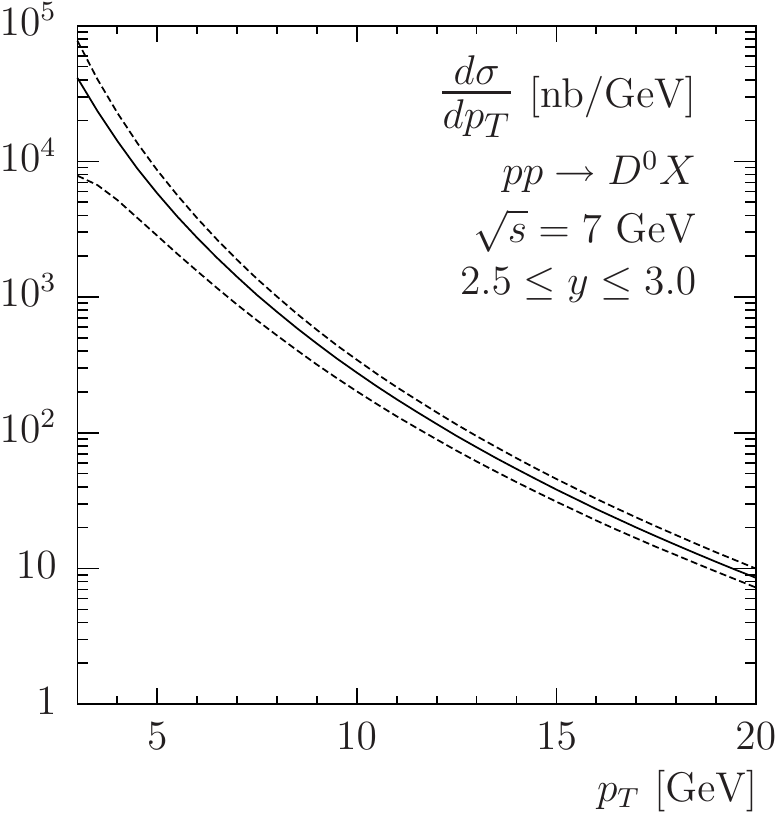}
\\[1.5ex]
\includegraphics[scale=0.72]{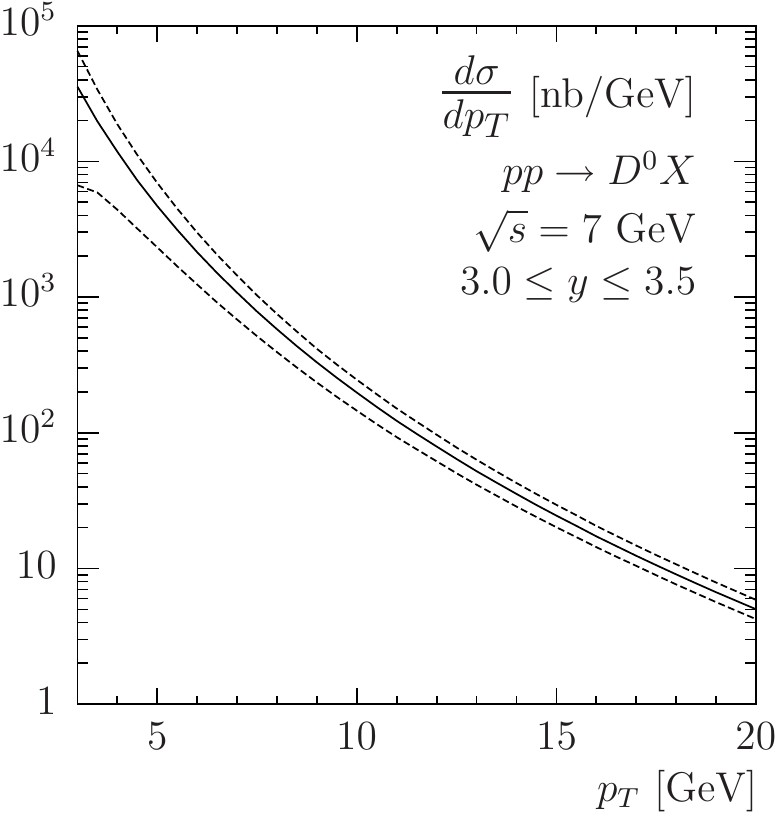}
~~~~~
\includegraphics[scale=0.72]{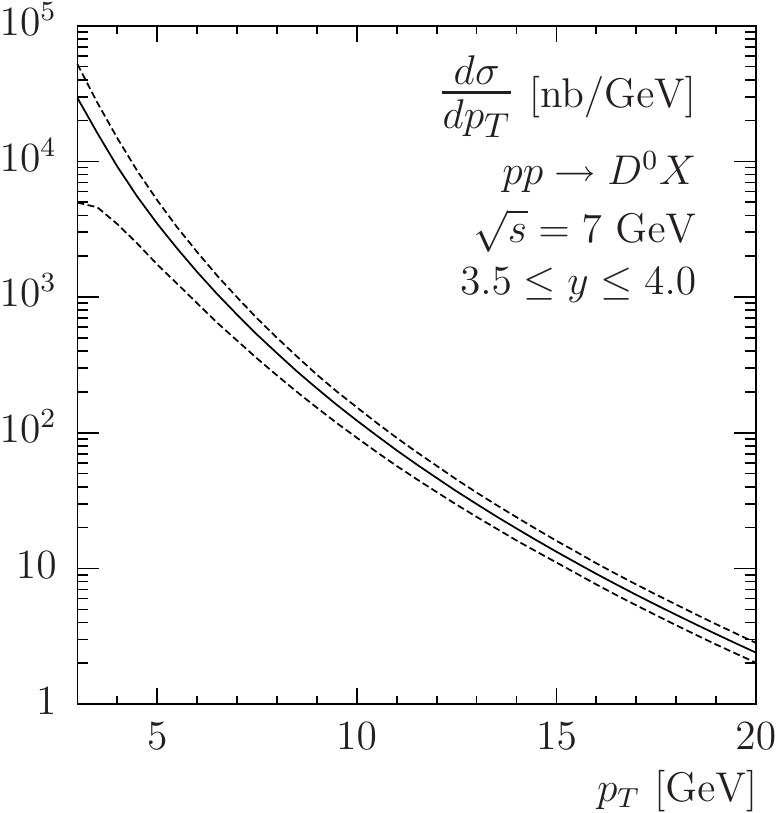}
\\[1.5ex]
\includegraphics[scale=0.72]{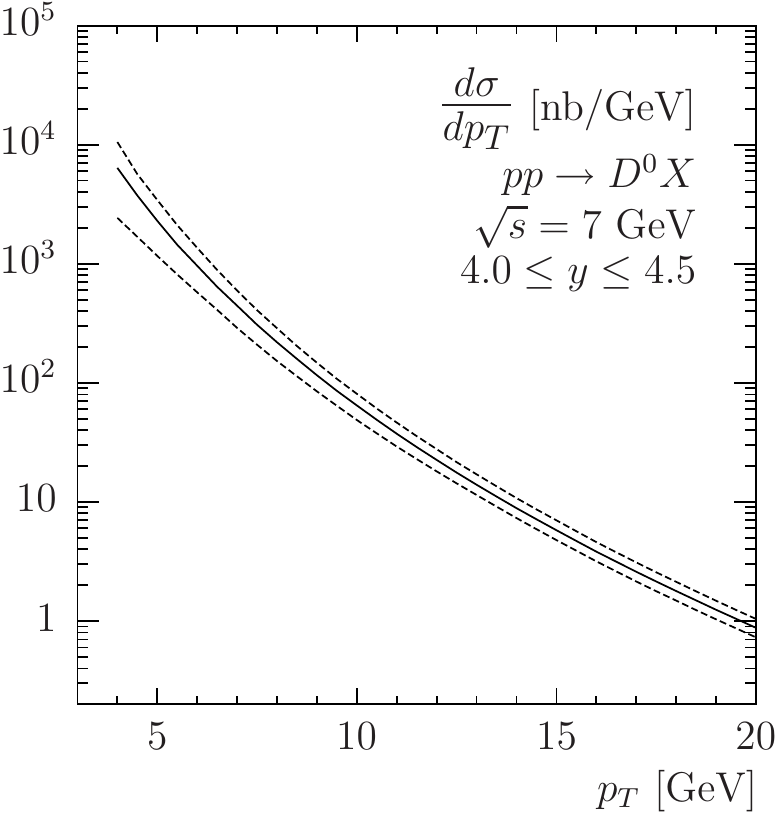}
~~~~~
\includegraphics[scale=0.72]{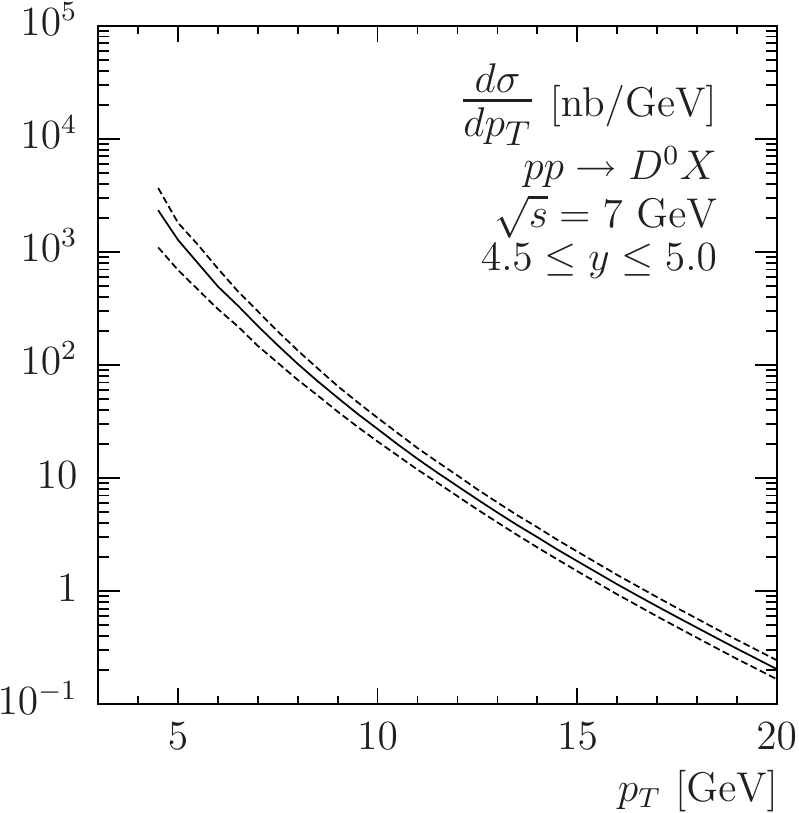}
\end{center}
\caption{\label{fig:9}
$p_T$ distributions $\mathrm{d}\sigma/\mathrm{d}p_T$ of $p+p \to 
D^0+X$ at $\sqrt{s}=7$~TeV in different rapidity bins as 
indicated in the figures. We show the uncertainty bands from scale 
variations following the prescription described in the text. The 
FFs are taken from Ref.~\cite{Kneesch:2007ey}.
} 
\end{figure}

The cross sections are largest at central rapidities and 
fall off towards larger values of $|y|$. This is shown in Fig.\ 
\ref{fig:7} for $D^0$, $D^+$, $D^{\ast}$ and $D_s$ mesons and 
in Fig.\ \ref{fig:8} for $\Lambda_c$ baryons. Here we present 
results for the $p_T$ distributions in six $|y|$ bins of width 
$\Delta y = 0.5$ between $y=2.0$ and $y=5.0$. Corresponding 
measurements made by the LHCb collaboration have been presented 
at conferences \cite{lhcb-confnote-2011}, and a publication is 
expected soon. At $p_T \simeq 3$ GeV, the cross section goes 
down with increasing rapidity by a factor of 3 to 5, depending 
on the type of the produced hadron; at $p_T \simeq 20$ GeV, 
the decrease with rapidity amounts to almost a factor of 50.
Again, uncertainties due to scale variations are large at 
small values of $p_T$ and decrease towards larger values of 
$p_T$. The corresponding results including also the theory 
error bands are shown in Fig.\ \ref{fig:9} for the case of 
$D^0$-meson production in the various rapidity bins. Within 
these errors, there is agreement with data as shown in Ref.\ 
\cite{lhcb-confnote-2011}.

\begin{figure}[b!]
\begin{center}
\begin{tabular}{cc}
\parbox{0.5\textwidth}{
\includegraphics[scale=0.53]{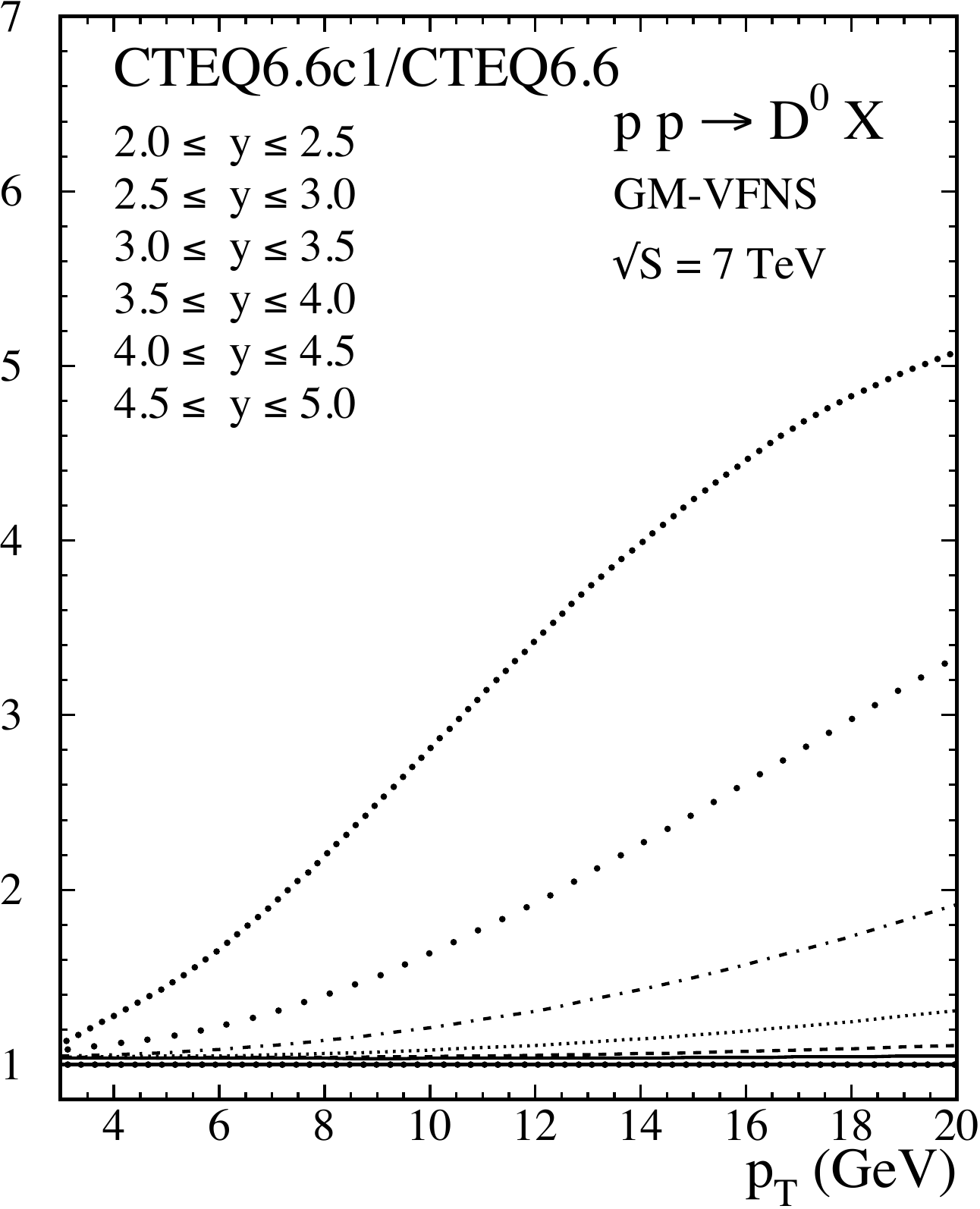}
}
&
\raisebox{-1.5mm}{\parbox{0.5\textwidth}{
\includegraphics[scale=0.53]{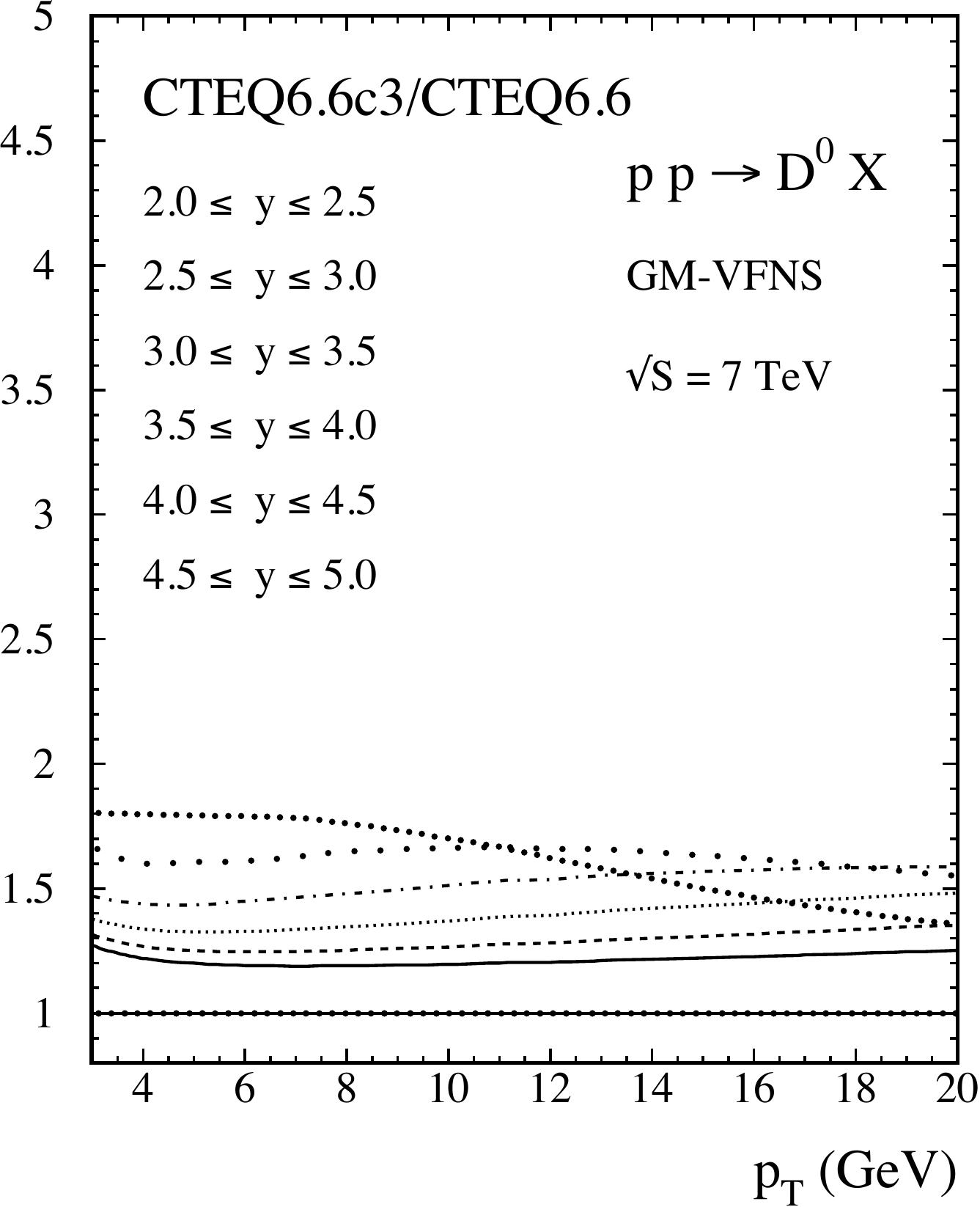}
}}
\\
(a) & (b) 
\end{tabular}
\end{center}
\caption{\label{fig:10}
Ratio of the $p_T$ distributions $\mathrm{d}\sigma/\mathrm{d}p_T$ 
for $p+p \to D^0 + X$ at NLO in the GM-VFNS, using different 
models of intrinsic charm: 
(a) BHPS model with 3.5\,\% $(c+\overline{c})$-content (at $\mu_F 
= 1.3$ GeV), 
(b) model with a high-strength sea-like charm component. 
The FFs are taken from Ref.\ \cite{Kneesch:2007ey} and $\sqrt{s} 
= 7$ TeV. The various lines represent the default predictions 
for $\xi_R = \xi_I = \xi_F = 1$, integrated over the rapidity 
regions indicated in the figures (larger rapidities correspond 
to larger cross section ratios everywhere in (a) and at small 
$p_T$ in (b)).
} 
\end{figure}

At large rapidities, it is interesting to study the influence 
of nonperturbative contributions to the charm-quark content 
of the incoming proton, usually called intrinsic charm. An 
enhancement of the charm PDF $c(x,\mu_F)$ at $x > 0.1$ can be 
visible in the cross section at large rapidities. The CTEQ 
collaboration has implemented appropriate models, which are 
compatible with the global data samples, in their PDF 
parametrization CTEQ6.5 \cite{Pumplin:2007wg}. In our recent 
work \cite{Kniehl:2009ar}, we studied the impact of these 
models on possible measurements at the Tevatron and at BNL RHIC. 
Here, we use the more recent parametrization CTEQ6.6 \cite{CTEQ6.6} 
and show corresponding results for the relative enhancements of 
the $p_T$ distributions in bins of rapidity in Fig.\ \ref{fig:10}. 
We have selected two models (see Ref.\ \cite{Pumplin:2007wg} for 
details): Fig.\ \ref{fig:10}a shows the calculation using the 
BHPS model with a 3.5\,\% $(c + \overline{c})$ content in the 
proton (at the scale $\mu_F = 1.3$ GeV), Fig.\ \ref{fig:10}b 
refers to the model of a high-strength sea-like charm component. 
We show results for $D^0$ production, but the cross section 
ratios for other $D$ mesons are very similar. One observes large 
enhancements, increasing with rapidity, and in the first model 
also with $p_T$. These numerical results show that it should be 
possible to exclude or narrow down models for intrinsic charm 
with forthcoming data from the LHCb experiment. 

\section{Conclusions}

In summary, we applied the GM-VFNS to obtain NLO predictions 
for the production of charmed mesons in $pp$ collisions at the 
LHC. Experimental data from the ALICE collaboration are already 
published, and agreement with our predictions is in general good, 
even at low values of $p_T$ if the factorization scale parameters 
are chosen appropriately. We expect more data from the other LHC 
experiments soon, and our results have been presented in a form 
which should make future comparisons straightforward. 

We have found that the production cross sections at large 
rapidities are sensitive to a nonperturbative component of 
the charm parton distribution function. Measurements should 
soon be able to exclude models which are still allowed by 
previous data.

\section*{Acknowledgments}

We thank our experimental colleagues from the LHC collaborations 
A.\ Dainese, P.\ Thompson, L.\ Gladilin, and M.\ Schmelling 
for discussions about the experimental results. 

This work was supported in part by the German Federal Ministry
for Education and Research BMBF through Grant No.\ 05~HT6GUA, 
by the German Research Foundation DFG through Grant No.\ 
KN~365/7--1, and by the Helmholtz Association HGF through 
Grant No.\ Ha~101.


\end{document}